# Earthquake tendency of the Himalayan seismic belt

Xue Lei[1,2], Qin Si-Qing[1,2,3]*, Yang Bai-Cun[1,2,3], Wu Xiao-Wa[1,2], Zhang Ke[1,2,3], Chen Hong-Ran[1,2,3]

[1] Key Laboratory of Shale Gas and Geoengineering, Institute of Geology and Geophysics, Chinese Academy of Sciences, Beijing 100029, P.R.China

[2] Institutions of Earth Science, Chinese Academy of Sciences, Beijing 100029, P.R.China

[3] University of Chinese Academy of Sciences, Beijing 100049, P.R.China

Corresponding author: Qin Si-Qing (E-mail: qsqhope@mail.iggcas.ac.cn)

**Abstract**

The theory about the brittle failures of multiple locked patches in a seismogenic fault system developed by us since 2010 is introduced in the present study. It is stated by the theory that the progressive failures of locked patch result in the occurrence of earthquakes due to fault movement, where the major earthquakes occurred at its volume expansion and peak strength points are referred to as characteristic ones. We analyze the seismogenic law of characteristic earthquakes in the Islamabad-Kathmandu seismic zone that extends along Himalayan seismic belt, and assess the seismicity trend of this zone by this theory. The results indicate that a $M_W$ 8.4 ~ 8.8 characteristic earthquake will strike the central Himalaya and the magnitude parameters of the 2015 Nepal earthquakes have a great impact on judging its occurrence time. When adopting the parameters by CEDC, the zone has almost reached the critical state and the expected characteristic one will take place in a short term (most likely within 10 years), probably with some $M_W \leq 7.0$ foreshocks. When adopting the parameters by NEIC, the occurrence time of the expected characteristic one cannot be judged at present until the critical state after the occurrence of some expected $M_W \leq 8.1$ preshocks is reached. Moreover, we reproduce the generation process of the 2015 Nepal earthquakes according to the self-similarity of seismogenic law for different-order locked patches.

**Keywords**: Himalayan seismic belt, 2015 Nepal earthquakes, Earthquake tendency, Earthquake prediction model,

Locked patch, Characteristic earthquake, Self-similarity of seismogenic law.

**1. Introduction**

The convergence between the India and Eurasian plates, as shown in Fig. 1, has triggered many great earthquakes along the Himalayan seismic belt (Bilham, 1995; Ambraseys and Bilham, 2000; Bilham and England, 2001; Ambraseys and Bilham, 2003; Ambraseys and Jackson, 2003; Ambraseys and Douglas, 2004; Bilham and Ambraseys, 2005; Rajendran and Rajendran, 2005; Wallace et al., 2005; Nayak et al., 2008; Rajendran et al., 2013; Sapkota et al., 2013; Srivastava et al., 2013; Gupta and Gahalaut, 2014; Patil et al., 2014), such as the 1505-06-06 $M$ ~8.2 Lo Mustang earthquake, the 1669-06-04 $M$ ~8.0 Pakistan-Rawalpindi earthquake, the 1803-09-01 $M$ ~8.1 Kumaon earthquake, the 1833-08-26 $M$ ~8.0 Nepal earthquake, the 1897-06-12 $M$ ~8.7 Shillong earthquake, the 1905-04-04 $M$ ~8.0 Kangra earthquake, the 1934-01-15 $M$ ~8.0 Nepal-Bihar earthquake and the 1950-08-15 $M$ ~8.6 China-Zayü earthquake. According to the National Earthquake Information Center (NEIC) and the China Earthquake Data Center (CEDC) (last accessed on 21 October 2016), a powerful earthquake of $M_W$ 7.8 by NEIC or $M_S$ 8.2 by CEDC struck Gorkha, Nepal on 25 April 2015 and it was followed by an earthquake of $M_W$ 7.3 by NEIC or $M_S$ 7.7 by CEDC occurred in Dolakha, Nepal on 12 May 2015, which resulted in serious casualties and property losses. In view of the occurrences of so many great earthquakes in the Himalayan seismic belt, undoubtedly it is a widespread concern for the issue whether a greater earthquake will hit this seismic belt in the future.

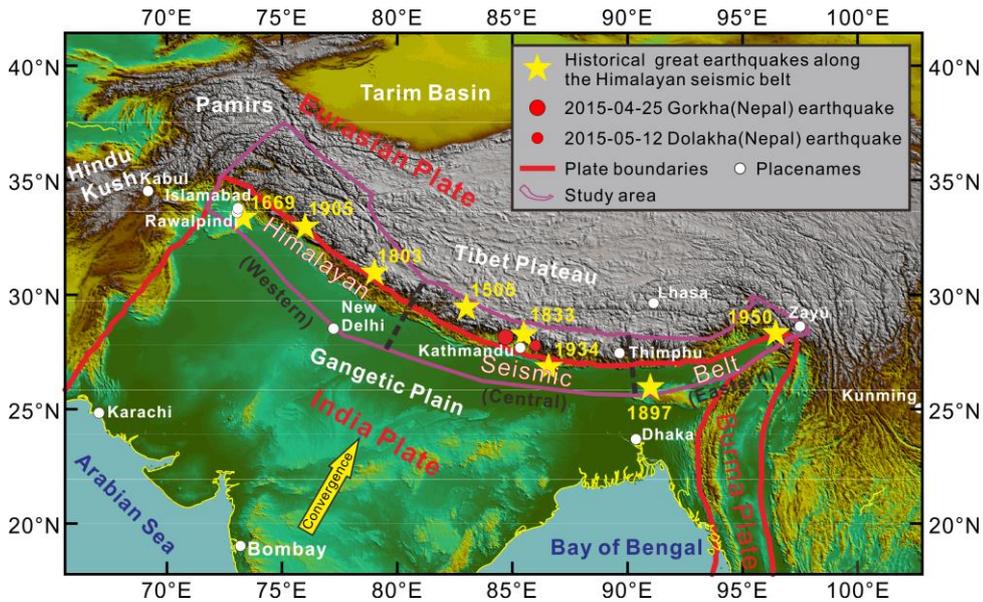

**Fig. 1 Seismotectonic map around the Himalayan seismic belt**, where the data of tectonic plate boundaries are from the United States Geological Survey (USGS) (http://earthquake.usgs.gov, last accessed on 11 September 2014) and two dashed black lines denote the boundaries among western, central and eastern segments of the Himalaya belt. The map was generated using ArcGIS 8.3.

Many detailed studies about the earthquake tendency in this region have been conducted since the occurrence of the 2015 Nepal earthquakes. For example, Feng et al. (2015) claimed that it is not unexpected that the region surrounding Kathmandu might experience more earthquakes in the future; Mitra et al. (2015) stated that other locked segments of the Himalaya could rupture in similar or greater earthquakes in the future; Zhang et al. (2015) suggested that Coulomb failure stress changes induced by the 2015 Gorkha earthquake could trigger new major earthquakes; Lindsey et al. (2015) found that there exists a 20 km slip gap between the two ruptures induced by the 2015 Gorkha and Dolakha earthquakes, and estimated that this unslipped fault patch has the potential to generate a $M_W$ 6.9 earthquake; Kobayashi et al. (2015) inferred that it is highly probable that a slip equivalent to $M_W$ ~7.0 would occur in the west side of the 2015 $M_W$ 7.3 Dolakha earthquake in the future; Mencin et al. (2016) reported that a future $M_W \leq 7.3$ earthquake could strike the unruptured region to the south and west of Kathmandu. Prakash et al. (2016) suggested that an unbroken portion of the plate boundary between the 1505 Lo Mustang earthquake and the 2015 Gorkha earthquake may generate an earthquake with a magnitude of about 8.0. Although these studies can offer some useful references for estimating the future seismic risk in the Himalayan seismic belt, there are still many unsolved problems on firm scientific grounds. For instance, are so many great earthquakes in this belt isolated or interrelated? Which kinds of events e.g. preshocks or foreshocks are the 2015 Nepal earthquakes? Can the 2015 Nepal earthquakes be predicted? How is the earthquake tendency of this seismic belt after the 2015 Nepal earthquakes?

Qin et al. (2010b) developed the theory about the brittle failures of multiple locked patches in a seismogenic fault system, which is closely associated with the process of accumulated damage leading to catastrophe for inhomogeneous media. Successful retrospective predictions of major or great earthquakes that occurred around the world state clearly that this theory can be verified repeatedly (Qin et al., 2016a; Qin et al., 2016b; Qin et al., 2016c; Wu et al., 2016). In view of this, we will analyze seismogenic law of great earthquakes occurred in the Himalayan seismic belt, assess the earthquake tendency of this seismic belt, and make enhanced insights into the above problems by the theory. Details of this theory will be introduced in Section 2.

## 2. Theory

### 2.1. Prediction model

Earthquakes result from rock fractures caused by fault movement. Qin et al. (2010b) believed that the fault's motion patterns and related seismic activity depend heavily upon one or multiple locked patches within faults. The locked patch associated with the spatial heterogeneity as illustrated in Fig. 2 can be defined as a high strength structural section within faults or subduction zones, including rock bridges, asperities and unruptured zones between discontinuous faults.

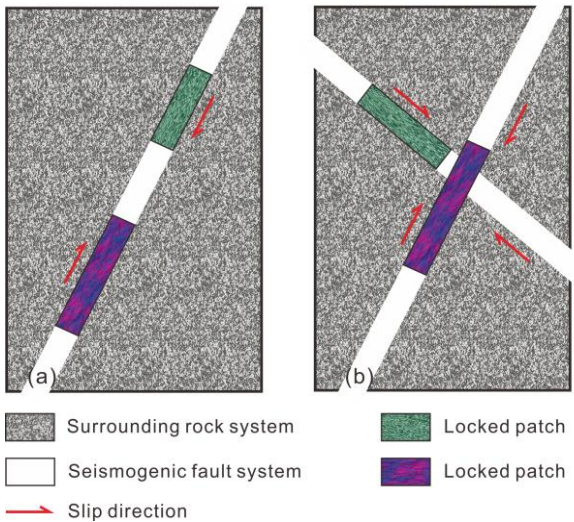

**Fig. 2 Schematic illustration of locked patches**. (**a**) A seismogenic fault includes two locked patches. (**b**) Each of two cross seismogenic faults has a locked patch.

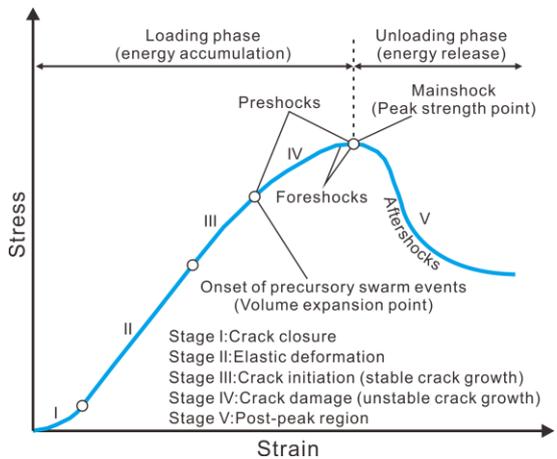

**Fig. 3 Schematic diagram illustrating different deformation stages of a rock specimen under the triaxial compression**.

Many studies (Brace et al., 1966; Bieniawski, 1967; Martin and Chandler, 1994; Qin et al., 2010b; Xue et al., 2014a; Xue et al., 2014b; Xue et al., 2015; Wu et al., 2016; Xue et al., 2017) showed that the deformation and failure process of rock specimens can be divided into five distinct stages (Fig. 3), among which the stable and unstable fracture stages are separated by the volume expansion point, while the unstable fracture stage and the post-peak failure stage are bounded by the peak strength point that is preceded by the volume expansion point for inhomogeneous media like rocks. It can be observed from rock mechanics test (Lei et al., 2004) that the spatial distributions of microseismic events in different evolution stages usually exhibit such an evolutionary pattern, from initially random to gradually concentrated with increasing load. The above analysis and interpretation based on the laboratory-scale rock failure process have laid a foundation for further revealing the fracture characteristics of large-scale natural locked patches, in view of the similarly hierarchical rupture behavior of different-scale rocks or seismic sources (Lei et al., 2003; Vallianatos et al., 2013; Yang et al., 2017a). Qin et al. (2010b) pointed out that when a single natural locked patch is regarded as a large-scale rock specimen, the occurrence of significant earthquakes (swarm events, see Fig. 4) or an accelerating strain increment from the onset of the volume expansion point of the locked patch is the only identifiable precursor to its macroscopic fracture, and that the first significant earthquake among the swarm events corresponds well to the volume expansion point of the locked patch. Clearly, the seismic events prior and posterior to the peak strength point of the locked patch represent that its energy is accumulated and released, respectively.

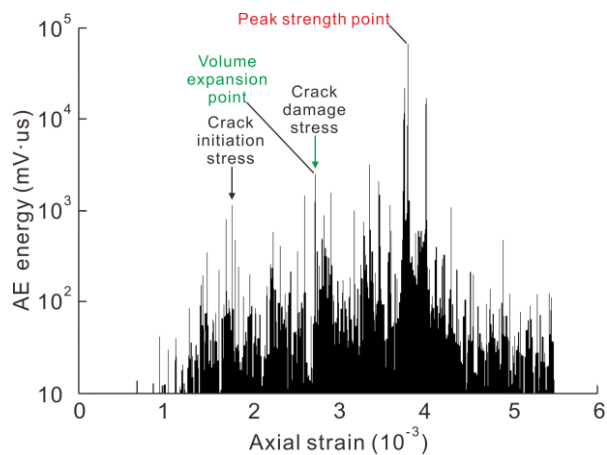

**Fig. 4 Variation characteristics of granite AE energy** (modified after Zhao et al. (2013)).

Starting from the above understanding and combining the re-normalization group theory with the damage constitutive model, Qin et al. (2010b) originally established a strain connection between the volume expansion point and the peak strength point of a single locked patch, which is especially suitable for describing the failure behavior of large-scale, slab-shaped rock specimens or locked patches (see Fig.2) subjected to an exceedingly slow rate of shear loading with a nearly fixed normal compressive stress or stress corrosion, i.e.,

$$\varepsilon_f = 1.48\varepsilon_c, \qquad (1)$$

where $\varepsilon_c$ and $\varepsilon_f$ are the shear strain values along a fault surface at the volume expansion and peak strength points of the locked patch, respectively.

Similarly, when involving multiple locked patches in a fault system, the following strain relation was further derived by Qin et al. (2010b), i.e.,

$$\varepsilon_f(k) = 1.48^k \varepsilon_c, \qquad (2)$$

where $\varepsilon_c$ and $\varepsilon_f(k)$ are the shear strain values corresponding to the volume expansion point of the first locked patch and to the peak strength point of the *k*-th locked patch, respectively. Eq. (2) has been verified by many retrospective cases e.g. laboratory-scale rock creep test (Yang et al., 2017a), rock avalanche and landslide with locked patches in the sliding surface (Qin et al., 2010a; Xue et al., 2014a; Chen et al., 2017). It is emphasized here that the constant 1.48 in Eq. (2) can help us to break through the obstacle of measuring the physical and mechanical parameters of rocks in deep earth, which makes it possible to predict major earthquakes.

Because the deep slip strain along a fault surface cannot be directly measured, we should seek an alternative physical quantity of strain. Fortunately, supposing that the fraction that stored elastic energy is converted into seismic waves is a constant for a defined seismic zone and the shear strain within a locked patch originated from sliding along a fault surface follows a uniform distribution, the cumulative Benioff strain (Benioff, 1949, 1951) can provide a measure of such a quantity (Qin et al., 2010b; Yang et al., 2017b). Thus, Eq. (2) can be written as

$$S_f(k) = 1.48^k S_c, \qquad (3)$$

where $S_c$ and $S_f(k)$ are the cumulative Benioff strain (CBS) values corresponding to the volume expansion point of the first locked patch and to the peak strength point of the *k*-th locked patch, respectively. Using the CBS value at the volume expansion point of the first locked patch, one can beforehand calculate the critical CBS value at the peak strength point of the *k*-th locked patch by Eq. (3) and prospectively judge the occurrence of major earthquakes or mainshocks.

Obviously, the premise of applying Eq. (3) to earthquake prediction is capable of extracting useful information of locked patch failures as much as possible from the earthquake catalog in a defined seismic zone which perhaps contains the seismic events not resulting from the failures of the locked patches themself. To this end, an approach to determine the minimum valid magnitude $M_v$ was presented by Qin et al. (2015) on the basis of the previous estimation method of the minimum completeness magnitude $M_c$ (Woessner and Wiemer, 2005). This approach has been successfully tested by case studies (Qin et al., 2016a; Qin et al., 2016b; Qin et al., 2016c; Wu et al., 2016) and it is pointed out that $M_v$ is usually not less than $M_c$.

Because the earthquake catalog prior to the volume expansion point of the first locked patch is usually still incomplete especially before 1900, even if $M_v$ or $M_c$ is considered, an initial error of CBS may arise. Therefore, we put forward an error correction expression derived from Eq. (3), i.e.,

$$\Delta = \frac{S_f^*(1) - 1.48 S_c^*}{0.48}, \tag{4}$$

where $\Delta$ denotes the error, $S_c^*$ and $S_f^*(1)$ are the uncorrected monitored values of CBS at the volume expansion and peak strength points of the first locked patch, respectively.

Here, we define the earthquakes occurred at the volume expansion and peak strength points as characteristic ones, earthquakes between these two points as preshocks as illustrated in Fig. 3, and the preshocks quite close to the peak strength point as foreshocks. Considering that the earlier the time is, the worse the completeness and accuracy of the event records in earthquake catalog are, especially for non-instrumental earthquake catalog, either earlier characteristic earthquakes were not recorded or the preshocks between the adjacent characteristic earthquakes were incompletely and (or) inaccurately recorded. Thus in data processing, we can only view the first significant earthquake whose corresponding value of CBS satisfies Eq. (3) after the error correction as the first characteristic earthquake occurred at the volume expansion point of the first locked patch.

Letting $M_C$ and $M_F$ denote the magnitude values of characteristic earthquakes corresponding to the volume expansion and peak strength points of locked patch, respectively, and $M_P$ denotes the magnitude value of preshocks or foreshocks, Wu et al. (2016) suggested that their magnitude relation in the condition of the same magnitude scale is generally constrained by

$$|M_F - M_C| \leq 0.5, \tag{5}$$

$$M_P \leq \min(M_F, M_C) - 0.2, \tag{6}$$

and that the magnitude values of several successive characteristic earthquakes usually follow such relation as

$$|M_F(k+1) - M_F(k)| \leq 0.2. \tag{7}$$

where $M_F(k)$ and $M_F(k+1)$ denote the magnitude values corresponding to the peak strength points of the $k$-th and $(k+1)$-th locked patches, respectively.

It is noted that Eqs. (5)-(7) are the constraint conditions of Eq. (3) and thus can be used to judge the rationality of magnitude values of significant earthquakes and to predict the magnitude values of characteristic earthquakes and preshocks.

**2.2. Seismic zoning method**

In addition to the prediction model mentioned above, an indispensable element for earthquake prediction is to define reasonably a seismic zone. As stated by Keilis-Borok (1994), the earthquake-prone areas in the earth can be divided into the hierarchical blocks with the relative movement by different-scale faults or fault networks, e.g. tectonic plates (Wilson, 1965, 1966; McKenzie and Parker, 1967; Le Pichon, 1968; Morgan, 1968), fault blocks (Zhang et al., 1979; Zhang et al., 1981a), and active tectonic blocks (Zhang et al., 2003; Zhang et al., 2005). In this context, we proposed that a seismic zone can be defined as an active-block region constrained by main regional faults or by plate boundaries. Considering that the strength of a fault is much lower than that of a block, tectonic deformations are mainly concentrated in the boundary faults, whereas blocks move essentially as a whole with a relatively smaller rate of internal deformations. Thus, we believed that the fault activities (earthquakes) within a defined-block region (i.e., seismic zone) are closely related to each other, while its adjacent-block regions only affect the loading or unloading mode of this specified-block region by shearing or extruding, but do not alter the intrinsic evolution law reflected by the internal seismicity of this specified-block region (Wu et al., 2016). Obviously, the seismic zoning concept proposed by us highlights the interaction among the seismogenic faults in a specified seismic zone, which is different from the conventional cognition that earthquakes occur along an individual fault or the rupture of a single fault is independent of one another. A latest finding of fault-jumping behavior observed for the 2016 $M_W$ 7.8 Kaikōura earthquake in New Zealand (Hamling et al., 2017; Mason, 2017) provides new evidence to support our viewpoint of seismic zoning. Relying on the data of faults and plate boundaries illustrated in the seismotectonic map (Zhang et al., 1981b; Li et al., 1997; Deng et al., 2007; Miao, 2010; Shen et al., 2014), we have defined 62 seismic zones around the world up to now (see Fig. 5), including 33 seismic zones in China and its adjacent areas (see Fig. 6), retrospectively analyzed the seismogenic law of characteristic earthquakes in these zones with satisfied results, and prospectively forecast the seismic trends in these zones (Qin et al., 2016a; Qin et al., 2016b; Qin et al., 2016c; Wu et al., 2016).

**Fig. 5 Division map of main seismic zones in the world (Version 2.0)**, where the data of tectonic plate boundaries are from USGS (http://earthquake.usgs.gov, last accessed on 11 September 2014). The map was generated using ArcGIS 8.3.

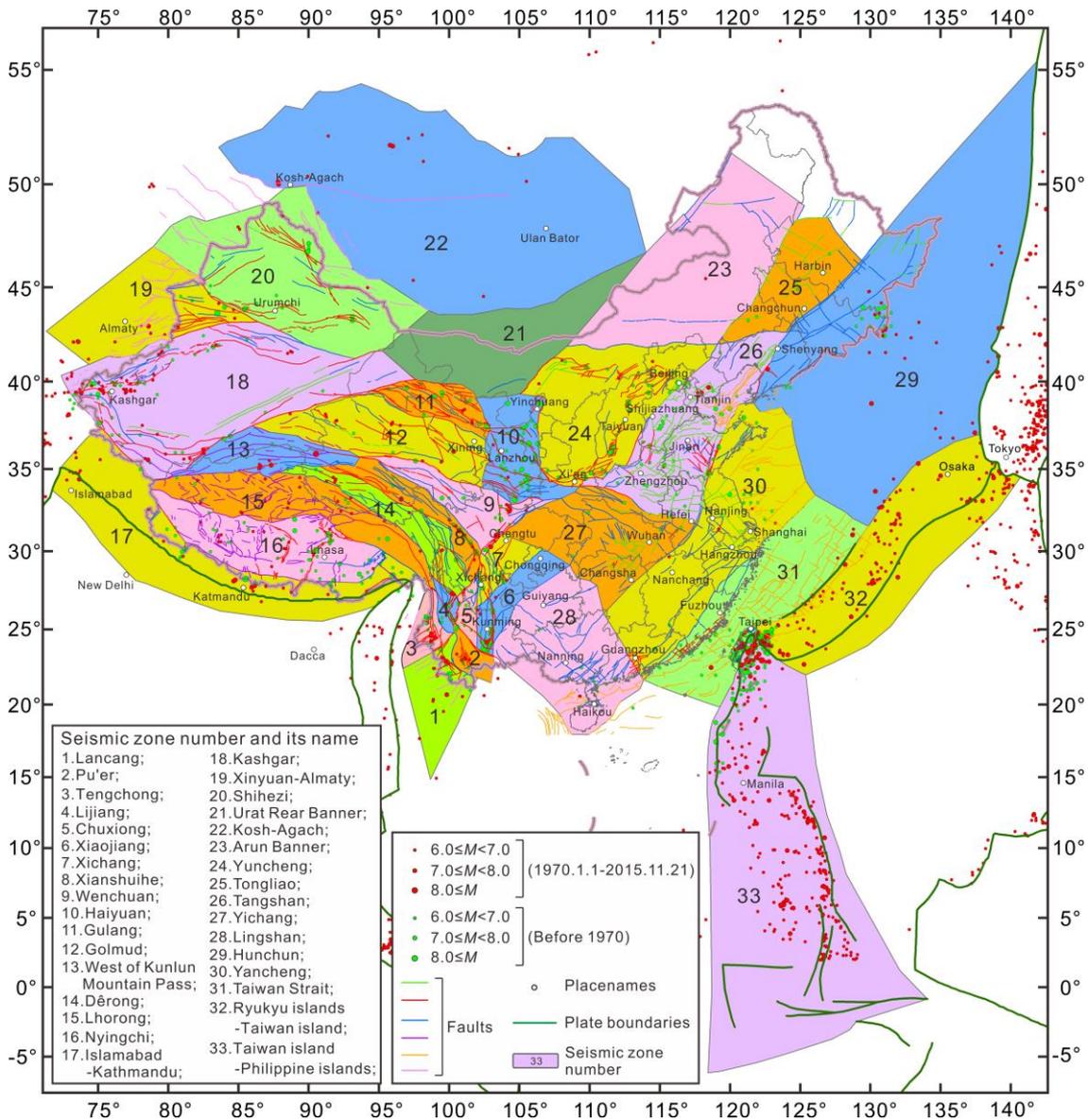

**Fig. 6 Division map of seismic zones in China and its adjacent areas (Version 3.6)**, where the fault data are from the Map of Active Tectonic in China (Deng et al., 2007) and the data of tectonic plate boundaries are from USGS (http://earthquake.usgs.gov, last accessed on 11 September 2014). The map was generated using ArcGIS 8.3.

It is emphasized here that the seismic zoning scheme's rationality must be examined by Eqs. (3)-(7). Taking the Tangshan seismic zone (No. 26 in Figs. 5 and 6) as an example, it is surrounded by seven seismic zones and constrained by some major regional faults as illustrated in Fig. 7, such as the Yangce-Gushi-Feizhong fault, Tieluzi-Luanchuan-Nanzhao fault, the Changzhi fault, the Chifeng-Kaiyuan fault, and the Tanlu fault belt including the Yishu fault and the Yilan-Yitong fault. Five characteristic earthquakes occurred in the current seismic period of the Tangshan seismic zone (Li et al., 2016), i.e., the 1597-10-06 $M_S$ 7.5 Bohai Sea earthquake, the 1668-07-25 $M_S$ 8.0 Tancheng earthquake, the 1679-09-02 $M_S$ 7.8 Sanhe-Pinggu earthquake, the 1888-06-13 $M_S$ 7.8 Bohai Sea earthquake and the 1976-07-27 $M_S$ 7.8 Tangshan earthquake. It can be seen from Fig. 8 that not only the seismogenic law of these characteristic earthquakes follows Eq. (3), but also the magnitude relation among $M_C$, $M_F$ and $M_P$ complies with Eqs. (5)-(7). Thus, we judge that the zoning scheme of the Tangshan seismic zone is reasonable. It should be pointed out that the 1830-06-12 $M_S$ 7.5 Cixian earthquake is a significant preshock before the 1888 Bohai Sea characteristic earthquake and the 1969-07-18 $M_S$ 7.4 Bohai Sea earthquake is a significant foreshock before the 1976 Tangshan characteristic earthquake.

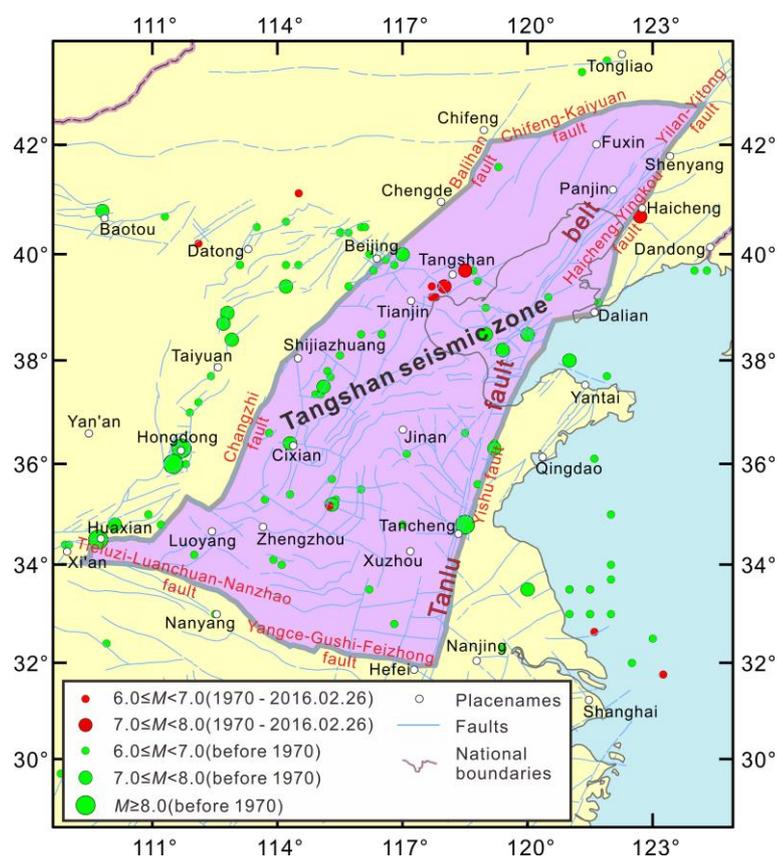

**Fig. 7 Seismotectonic map of the Tangshan seismic zone**, where the fault data are from the Map of Active Tectonic in China (Deng et al., 2007). The map was generated using ArcGIS 8.3.

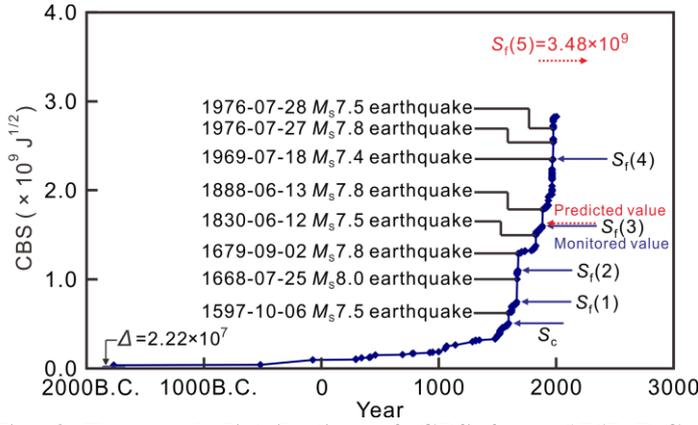

**Fig. 8 Temporal distribution of CBS from 1767 B.C. to 2015-11-21 for the Tangshan seismic zone.** Earthquakes with $M_S \geq 5.0$ are selected for data analysis, i.e., $M_v = M_S$ 5.0. The error $\Delta$ is calculated by Eq. (4).

**2.3. Identification method of mainshock**

Another indispensable element for earthquake prediction is to judge whether a characteristic earthquake is a mainshock or not. The mainshock in a seismic period, as pointed out by Qin et al. (2010b), will inevitably take place in a defined seismic zone when the last locked patch is loaded to reach its macroscopic fracture point, i.e., the peak strength point. After the mainshock and its subsequent aftershocks, a new cycle of seismic period will begin again (see Fig. 9). Obviously, correctly identifying the mainshock is crucial to judge the seismic trend of a defined seismic zone.

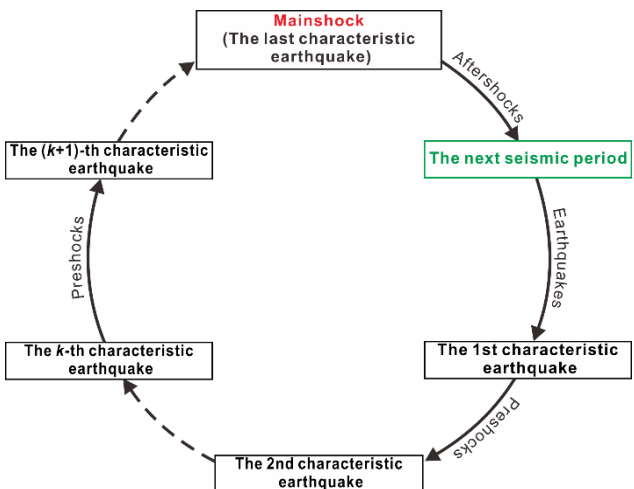

**Fig. 9 Cycle of seismic period for a defined seismic zone.**

After the occurrence of a characteristic earthquake corresponding to the peak strength point of locked patch in a defined seismic zone, if the following relation is satisfied, i.e.,

$$E_C \ll E_B, \tag{8}$$

where $E_B$ denotes the accumulated elastic strain energy of the seismic zone prior to this characteristic earthquake, and $E_C$ denotes the released energy of itself, we can judge that it is not a mainshock in the current seismic period, based on the conservation law of elastic strain energy (Wu et al., 2016), i.e.,

$$E_T = E_M + E_A, \tag{9}$$

where $E_T$ is the accumulated elastic strain energy of the seismic zone prior to the mainshock, $E_M$ is the released elastic strain energy of the mainshock itself, and $E_A$ is the total elastic strain energy released by aftershocks.

For example, a mainshock, the 1515-06-27 $M_S$ 7.75 Yongsheng earthquake in Yunnan, China, took place in the last seismic period of the Lijiang seismic zone (Qin et al., 2016b) (No. 4 in Figs. 5 and 6), but the mainshock has not yet occurred in the current seismic period because these three major earthquakes, i.e., the 1751-05-25 $M_S$ 6.75 Jianchuan earthquake, the 1925-03-16 $M_S$ 7.1 Dali earthquake and the 1996-02-03 $M_S$ 6.9 Lijiang earthquake in Yunnan, are characteristic ones but not mainshocks by Eq. (8) (see Fig. 10). We can infer that the magnitude value of the mainshock in the current seismic period is also about $M_S$ 7.75.

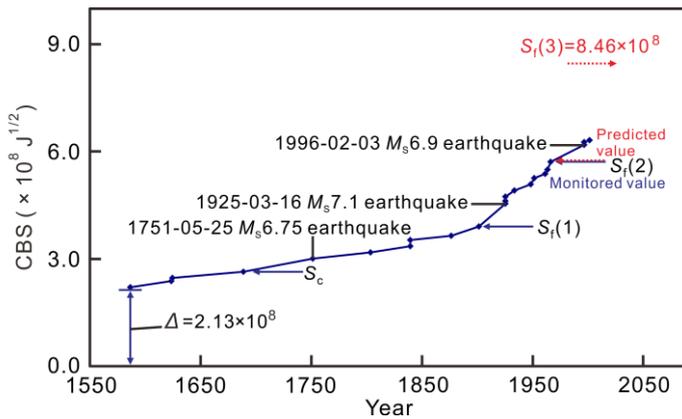

**Fig. 10 Temporal distribution of CBS from 1586-04-02 to 2015-11-21 for the Lijiang seismic zone**. Earthquakes with $M_L \geq 5.9$ are selected for data analysis, i.e., $M_v = M_L$ 5.9. The error $\Delta$ is calculated by Eq. (4).

As another example, three characteristic earthquakes occurred in the current seismic period of the Hokkaido seismic zone (Wu et al., 2016) (No. 39 in Fig. 5), i.e., the 1898-06-05 $M_{uk}$ 8.7 Honshu earthquake in Japan where the magnitude scale $M_{uk}$ represents that its related calculation method or published source is not clearly determined and is estimated by the statistical method (Song et al., 2011), the 1952-11-04 $M_W$ 8.9 Kamchatka earthquake and the 2011-03-11 $M_W$ 9.0 Tohoku earthquake in Japan. It can be seen from Fig. 11 that the seismogenic law of these characteristic earthquakes follows Eq. (3). One can judge by Eq. (8) that both the 1952 Kamchatka earthquake and 2011 Tohoku earthquake are not mainshocks. This implies that a characteristic earthquake with a magnitude of $M_W$ 9.0 ~ 9.5 will hit this seismic zone again when the critical CBS value calculated by Eq. (3), $4.79 \times 10^{10}$ J$^{1/2}$, is reached. Similarly, it can also be judged by Eq. (8) that characteristic earthquakes occurred in the Jakarta seismic zone (No. 34 in Fig. 5), e.g. the 1818-11-08 $M_S$ 8.5 Indonesia Bali Sea earthquake, the 1861-02-16 $M_S$ 8.5 Lagundi earthquake, the 1938-02-01 $M_W$ 8.5 Banda Sea earthquake and the 2004-12-26 $M_W$ 9.0 Sumatra earthquake, are not mainshocks (Wu et al., 2016), implying that a characteristic earthquake with a magnitude of $M_W$ 9.0 ~ 9.5 will strike this seismic zone when the critical CBS value, $1.99 \times 10^{10}$ J$^{1/2}$, is reached (see Fig. 12).

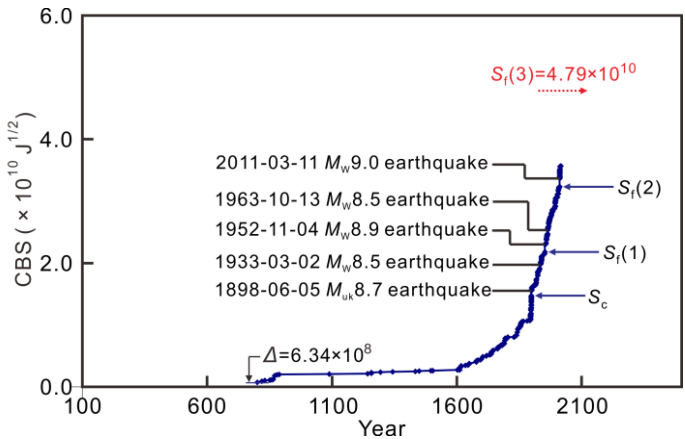

**Fig. 11 Temporal distribution of CBS from 144-02-15 to 2016-02-24 for the Hokkaido seismic zone**. Earthquakes with $M_W \geq 7.0$ are selected for data analysis, i.e., $M_v = M_W$ 7.0. The error $\Delta$ is calculated by Eq. (4).

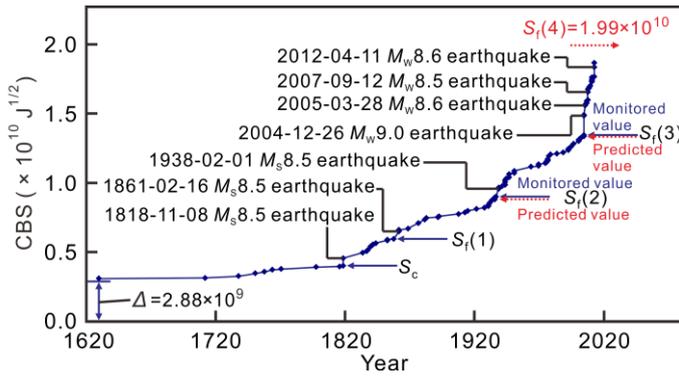

**Fig. 12 Temporal distribution of CBS from 1629-08-01 to 2016-02-24 for the Jakarta seismic zone.** Earthquakes with $M_S \geq 7.0$ are selected for data analysis, i.e., $M_v = M_S$ 7.0. The error $\Delta$ is calculated by Eq. (4).

## 3. Earthquake tendency of study area

### 3.1. Overview of study area

The study area (Fig. 1) extends along the Himalayan seismic belt and is referred to as the Islamabad-Kathmandu seismic zone (No. 17 in Figs. 5 and 6), which is surrounded by seven seismic zones and constrained by some major regional faults, such as the Karakoram fault and the South Tibet Detachment System.

### 3.2. Data source and data processing method

In the current study, the pre- and post- 1900 earthquake catalogs are cited from the "Global Earthquake Catalog" complied by Song et al. (2011) and NEIC (last downloaded on 16 February 2016), respectively. According to the editorial comments of "Global Earthquake Catalog", it can be learnt that this book is compiled through collecting and revising earthquake catalogs from different websites and publications in the world, and gives a priority to seismic parameters provided by (Engdahl and Villaseñor, 2002) when there exist different access channels for a certain earthquake. As reported in our previous studies (Qin et al., 2016a; Qin et al., 2016b; Qin et al., 2016c; Wu et al., 2016), most of the earthquake catalog provided in this book is with high accuracy.

In data processing, different magnitude scales should be first uniformly converted into the local magnitude $M_L$ using the conversion equations suggested by Yang et al. (2017b), and then seismic moment, seismic energy, Benioff strain, and CBS are computed consecutively.

### 3.3. Seismogenic law of historical characteristic earthquakes

Seven $M ≥ 8.0$ earthquakes occurred in the Islamabad-Kathmandu seismic zone since 1505 are listed in Table 1, where the magnitude scale $M_K$ is defined and estimated by the level of destruction (Song et al., 2011) and is regarded as the local magnitude $M_L$ in the data processing of present study. It is noted that the magnitude parameters of the 1897 Shillong earthquake given by some authors are quite different from each other, such as $M_W$ 8.1 (Bilham and England, 2001) and $M_S$ 8.7 (Richter, 1958; Song et al., 2011). We think that the magnitude parameter $M_S$ 8.5 is reasonable and adopted in this study because of the following reasons. Firstly, taking into account the 1833 Nepal earthquake with a magnitude of $M_S$ 8.0, the magnitude value of the Shillong earthquake should not exceed $M_S$ 8.5 by Eq. (5). Secondly, due to the 1950 China-Zayü earthquake with a magnitude of $M_W$ 8.6, the magnitude value of the Shillong earthquake is most likely to be more than $M_S$ 8.4 by Eq. (7). Also, the 1934-01-15 Nepal-Bihar earthquake is given as $M_W$ 8.0 by NEIC, $M_S$ 8.2 by ISC (International Seismological Centre; last accessed on 8 October 2016), and $M_W$ 8.0 ± 0.3 by the ISC-GEM Global Instrumental Earthquake Catalog (http://www.isc.ac.uk/iscgem/ ; last accessed on 8 October 2016), respectively. In view of the 1897 $M_S$ 8.5 Shillong earthquake, we judge by Eq. (6) that the magnitude value of the 1934 Nepal-Bihar earthquake should not exceed $M_S$ 8.3 and the parameter $M_S$ 8.2 given by ISC is reasonable, which is adopted in this study.

**Table 1. The earthquake events with $M ≥ 8.0$ in the Islamabad-Kathmandu seismic zone**

| No. | Date | Latitude/Longitude (°) | Depth (km) | Magnitude | Magnitude scale | Source |
| --- | --- | --- | --- | --- | --- | --- |
| 1 | 1505-06-06 | 29.50/83.00 | | 8.2 | $M_S$ | Song et al. (2011) |
| 2 | 1669-06-04 | 33.40/73.30 | | 8.0 | $M_K$ | Song et al. (2011) |
| 3 | 1803-09-01 | 31.00/79.00 | | 8.1 | $M_S$ | Song et al. (2011) |
| 4 | 1833-08-26 | 28.30/85.50 | | 8.0 | $M_S$ | Song et al. (2011) |
| 5 | 1897-06-12 | 26.00/91.00 | 33.0 | 8.7 | $M_S$ | Song et al. (2011) |
| 6 | 1934-01-15 | 26.89/86.59 | 15.0 | 8.0 | $M_W$ | NEIC |
| 7 | 1950-08-15 | 28.36/96.45 | 15.0 | 8.6 | $M_W$ | NEIC |

By applying the magnitude parameters of 2015 Nepal earthquakes measured respectively by CEDC and NEIC, the mechanical correlation among historical characteristic earthquakes occurred in the current seismic period of the Islamabad-Kathmandu seismic zone is shown in Fig. 13 after the error correction. According to the CBS value prior to the 1669-06-04 $M_K$ 8.0 Pakistan-Rawalpindi earthquake and using Eq. (3), one can accurately and continuously forecast the critical CBS values of the 1803-09-01 $M_S$ 8.1 Kumaon earthquake, the 1833-08-26 $M_S$ 8.0 Nepal earthquake, the 1897-06-12 $M_S$ 8.5 Shillong earthquake and the 1950-08-15 $M_W$ 8.6 China-Zayü earthquake. It is noted that the 1934 $M_S$ 8.2 Nepal-Bihar earthquake is a significant preshock occurred between the 1897 Shillong earthquake and the 1950 China-Zayü earthquake. The satisfied retrospective analysis demonstrates the reliability of the theory described in Section 2, which provides a firm basis for prospectively predicting the occurrence of the next characteristic earthquake and assessing the earthquake tendency of this seismic zone.

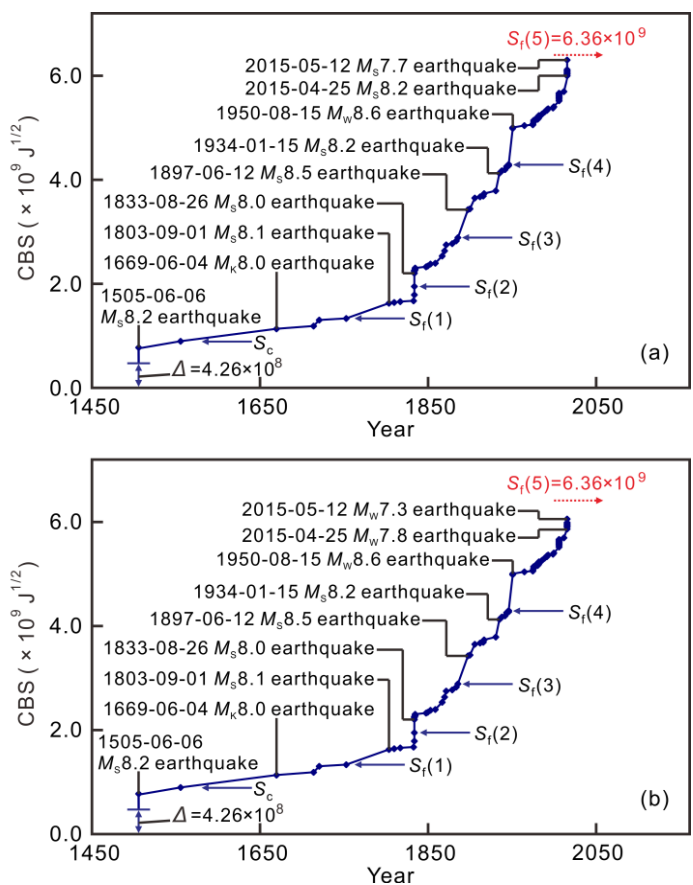

**Fig. 13 Temporal distribution of CBS from 1505-06-06 to 2016-02-16 for the Islamabad-Kathmandu seismic zone.** Earthquakes with $M_L \geq 6.4$ are selected for data analysis, i.e., $M_v = M_L$ 6.4. The error $\Delta$ is calculated by Eq.

(4). (**a**) The magnitude parameters of 2015 Nepal earthquakes by CEDC. (**b**) The magnitude parameters of 2015 Nepal earthquakes by NEIC.

### 3.4. Earthquake tendency

We judge by Eq. (8) that the 1950 $M_W$ 8.6 China-Zayü earthquake is not a mainshock, because the accumulated elastic strain energy (about $1.03 \times 10^{18}$ J) of this seismic zone prior to it is far more than the elastic strain energy (about $4.89 \times 10^{17}$ J) released by itself, implying that a characteristic earthquake will strike the Islamabad-Kathmandu seismic zone again when the critical CBS value, $6.36 \times 10^9$ J$^{1/2}$, is reached. The time-space-magnitude evolution of the expected characteristic earthquake as well as the preshocks or foreshocks prior to it is analyzed as follows.

### 3.4.1. Magnitude

Because of the 1950 China-Zayü earthquake with a magnitude of $M_W$ 8.6, the magnitude value of the expected characteristic earthquake is estimated to be $M_W$ 8.4 ~ 8.8 by Eq. (7).

As mentioned above, when adopting the magnitude parameters of the 2015 Nepal earthquakes by CEDC, the Islamabad-Kathmandu seismic zone has almost reached the critical state. We infer that some $M_W \leq 7.0$ foreshocks may occur prior to the expected characteristic earthquake. When adopting the magnitude parameters of the 2015 Nepal earthquakes by NEIC, the difference between the predicted critical value and monitored one of CBS is about $3.0 \times 10^8$ J$^{1/2}$, which is approximately equivalent to the Benioff strain corresponding to a $M_W$ 8.1 earthquake. In other words, the maximum magnitude value of preshocks or foreshocks prior to the expected characteristic earthquake will not exceed $M_W$ 8.1.

### 3.4.2. Occurrence time

When the magnitude parameters of the 2015 Nepal earthquakes by CEDC is adopted, the monitored value of CBS at present is about $6.33 \times 10^9$ J$^{1/2}$, almost equal to the critical one, $6.36 \times 10^9$ J$^{1/2}$ (Fig. 13a). In other words, the 2015 Nepal earthquakes can be regarded as significant foreshocks, implying that the Islamabad-Kathmandu seismic zone has almost located at the critical state. As reported by Li et al. (2016), when a seismic zone reaches or quite approaches the critical state after foreshocks, the expected characteristic earthquake will take place after a delay of a few years to decades, which may be attributed to the viscosity effect of the seismogenic fault. For example, the seismic zone of the west of Kunlun mountain pass (No. 13 in Figs. 5 and 6) was located at the critical state after the 1997-11-08 $M_S$ 7.3 Mani foreshock in Tibet, China (Qin et al., 2016d) (see Fig. 14), but the expected characteristic earthquake, the 2001-11-14 $M_S$ 8.0 Kokoxili earthquake in Tibet, China, did not take place until about 4 years later. For another example, the Northern Mariana Islands seismic zone (No. 38 in Fig. 5) was located at the critical state since the occurrence of 2010-12-21 $M_W$ 7.4 Bonin Islands foreshock (Qin et al., 2016a), but the expected characteristic earthquake, the 2015-05-30 $M_W$ 7.9 Bonin Islands earthquake, did not take place until about 5 years later (see Fig. 15). By analogy to the above earthquake cases and considering that the delay time of the previous characteristic earthquakes in the Islamabad-Kathmandu seismic zone, such as the 1897 $M_S$ 8.5 Shillong earthquake and the 1950 $M_W$ 8.6 China-Zayü earthquake, is about 12 years and 7 years after foreshocks, respectively, we deduce that the expected characteristic earthquake will take place in a short term, most likely within 10 years since 2017.

When the magnitude parameter by NEIC is adopted, the monitored value of CBS in this seismic zone up to 26 February 2016 is about $6.06 \times 10^9$ $J^{1/2}$, less than the critical one, $6.36 \times 10^9$ $J^{1/2}$ (Fig. 13b). Therefore, the occurrence time of the expected characteristic earthquake cannot be judged at present until the critical state after the occurrence of some expected $M_W \leq 8.1$ preshocks is reached.

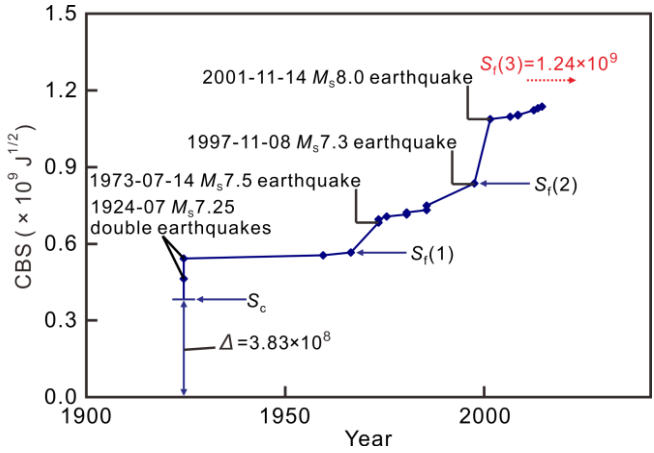

**Fig. 14 Temporal distribution of CBS from 1924-07-03 to 2015-11-21 for the seismic zone of the west of Kunlun mountain pass**. Earthquakes with $M_S \geq 5.6$ are selected for data analysis, i.e., $M_v = M_S$ 5.6. The error $\Delta$ is calculated by Eq. (4).

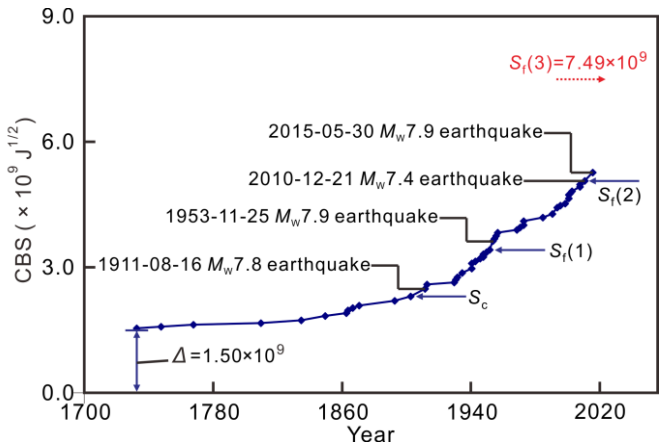

**Fig. 15 Temporal distribution of CBS from 1606-01-23 to 2016-02-24 for the Northern Mariana Islands seismic zone**. Earthquakes with $M_L \geq 7.0$ are selected for data analysis, i.e., $M_v = M_L$ 7.0. The error $\Delta$ is calculated by Eq. (4).

### 3.4.3. Potential earthquake source

According to the spatial distribution of unruptured segments along the Himalayan seismic belt (Bilham et al., 2001) and the current seismic activity, it is inferred that the expected characteristic earthquake is most likely to strike the central Himalaya (see Fig. 1). We will track the seismicity of the Islamabad-Kathmandu seismic zone for obtaining a better result.

### 3.5. Seismogenic law of the 2015 Nepal earthquakes

Understandably, when a large-scale first-order locked patch contains small-scale second-order locked patches due to the heterogeneity (see Fig. 16), the seismogenic law of characteristic preshocks corresponding to the volume expansion and peak strength points of second-order locked patches, as pointed out by Qin et al. (2016b) and Yang et al. (2017a, c), should also be governed by Eq. (3). Here, we define the earthquakes between the successive characteristic preshocks as sub-preshocks. Similarly, letting $M_{PC}$ and $M_{PF}$ denote the magnitude values of characteristic preshocks, respectively, and $M_{PP}$ denote the magnitude value of sub-preshocks, our preliminary research suggests that their magnitude relation in the condition of the same magnitude scale is generally constrained by

$$\left|M_{PF} - M_{PC}\right| \leq 1.0, \tag{10}$$

$$M_{PP} \leq \min(M_{PF}, M_{PC}). \tag{11}$$

Thus, Eqs. (10) and (11) can be used to estimate the magnitude values of characteristic preshocks and sub-preshocks.

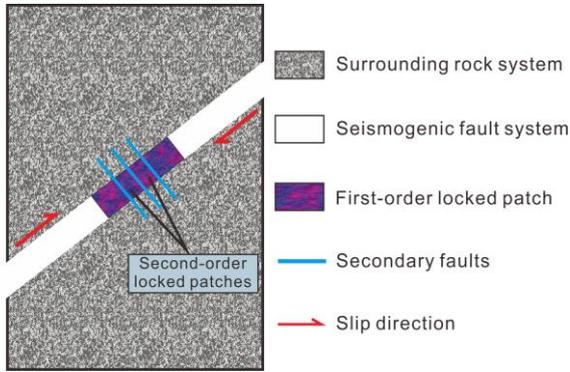

**Fig. 16 Schematic diagram illustrating the hierarchical structure of locked patch.**

Relying on the above-mentioned self-similarity of seismogenic law, the mechanical correlation among characteristic preshocks between the 1950-08-15 $M_W$ 8.6 China-Zayü characteristic earthquake and the expected $M_W$ 8.4 ~ 8.8 characteristic earthquake can be revealed after the error correction (see Fig. 17). According to the CBS value prior to the 1975-01-19 $M_S$ 6.8 Kinnaur preshock and using Eq. (3), one can more accurately and continuously forecast the critical CBS values of the 1981-09-12 mb 6.2 ($M_L$ 7.0) Kashmir preshock, the 2005-10-08 $M_W$ 7.6 Kashmir preshock and the 2015-04-25 Gorkha preshock. Thus, the 2015 Gorkha earthquake can be regarded as a characteristic preshock. The satisfied retrospective analysis confirms the self-similar seismogenic law for different-order locked patches, and further demonstrates the seismic zoning scheme's rationality of the Islamabad-Kathmandu seismic zone.

It can also be seen from Fig. 17 that the magnitude parameters of the 2015 Nepal earthquakes have a great impact on the predictability of the 2015-05-12 Dolakha earthquake. When applying the magnitude parameters of 2015 Nepal earthquakes measured by CEDC, the 2015 Dolakha earthquake can also be predicted and is also a characteristic preshock (see Fig. 17a). When applying the magnitude parameters of 2015 Nepal earthquakes measured by NEIC, the 2015 Dolakha earthquake cannot be predicted and is not a characteristic preshock (see Fig. 17b).

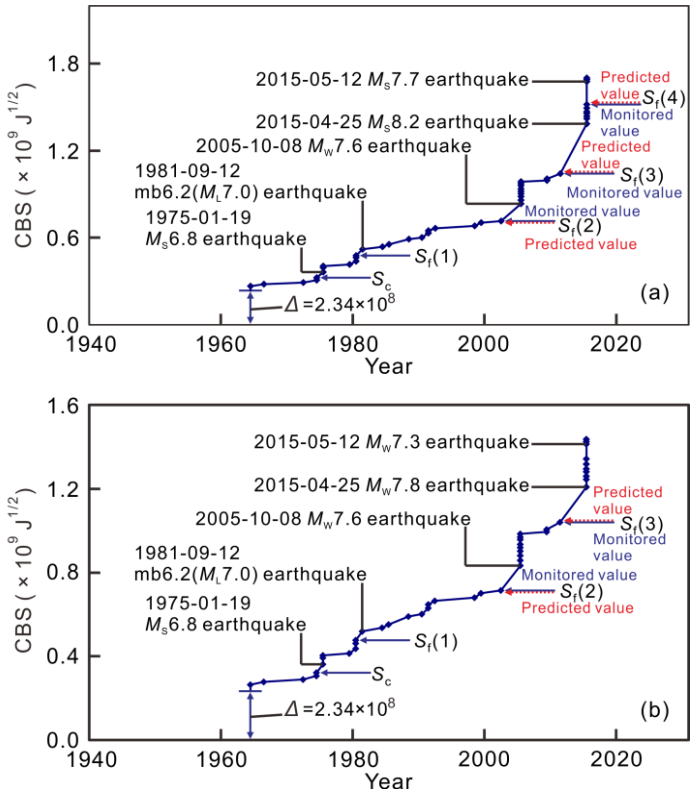

**Fig. 17 Temporal distribution of CBS from 1964-10-21 to 2016-02-16 for the Islamabad-Kathmandu seismic zone.** Earthquakes with $M_L \geq 6.1$ are selected for data analysis, i.e., $M_v = M_L$ 6.1. The error $\Delta$ is calculated by Eq. (4). (**a**) The magnitude parameters of 2015 Nepal earthquakes by CEDC. (**b**) The magnitude parameters of 2015

Nepal earthquakes by NEIC.

## 4. Conclusions

According to the theory about the brittle failures of multiple locked patches in a seismogenic fault system, we analyze the seismogenic law of historical characteristic earthquakes in the Islamabad-Kathmandu seismic zone, and conclude that a $M_W$ 8.4 ~ 8.8 characteristic earthquake will strike this seismic zone in the future. Obviously, it can be learnt from the present study that these characteristic ones along Himalayan seismic belt are interrelated, and both the 2015-04-25 Gorkha earthquake and 2015-05-12 Dolakha earthquake are preshocks or foreshocks of the next expected characteristic earthquake. When adopting the magnitude parameters of the 2015 Nepal earthquakes by CEDC, the expected characteristic one will take place in a short term (most likely within 10 years). When adopting the magnitude parameters by NEIC, the occurrence time of the expected characteristic earthquake cannot be judged at present. No matter what the situation is, the expected characteristic one will be most likely to strike the central segment of the Himalayan seismic belt based on the current seismic activity and the spatial distribution of unruptured segments in the Islamabad-Kathmandu seismic zone. Thus, we suggest that the countries concerned should pay more attention to the central segment of the Himalayan seismic belt and take effective measures for mitigating earthquake disasters.

In addition, at least the 2015-04-25 Gorkha earthquake can also be predicted by the theory, owing to the self-similarity of seismogenic law for the first-order and second-order locked patches.

**Acknowledgements**

This work was supported by the National Natural Science Foundation of China under grant numbers 41572311 and 41302233. We thank Prof. J.Y. Dong for his help in creating Figs. 1, 5, 6 and 7.

triaxial compression conditions. *Proceedings of the 3rd ISRM SINOROCK Symposium*. London: CRC Press, 75-80.